\documentclass[%
 reprint,
superscriptaddress,
 amsmath,amssymb,
 aps,
]{revtex4-2}

\usepackage{graphicx}
\usepackage{dcolumn}
\usepackage{float}
\usepackage{bm}
\usepackage[utf8]{inputenc} 
\usepackage{CJKutf8} 
\usepackage{xcolor} 
\usepackage{makecell}
\usepackage{amssymb}   
\usepackage[normalem]{ulem}

\def\frontmatter@thefootnote{\arabic{footnote}}
\begin{document}

\preprint{APS/123-QEC}

\title{An Architecture for Real-Time Quantum Error Correction with Neutral Atoms}
\author{X. Wang}
 \affiliation {Shanghai Zhongqi Wuliang Quantum Technology Co. Ltd., Shanghai, China}
\author{Y. Jing}
 \affiliation {Shanghai Zhongqi Wuliang Quantum Technology Co. Ltd., Shanghai, China}
\author{Z. Yu}
 \affiliation {The 32nd Research Institute of China Electronics Technology Corporation, Shanghai, China}
\author{S. Wang}
 \affiliation {The 32nd Research Institute of China Electronics Technology Corporation, Shanghai, China}
\author{D. Niu}
 \affiliation {Shanghai Zhongqi Wuliang Quantum Technology Co. Ltd., Shanghai, China}
\author{Y. Liu}
 \affiliation {Shanghai Zhongqi Wuliang Quantum Technology Co. Ltd., Shanghai, China}
\author{T. Xiao}
 \affiliation {State Key Laboratory of Photonics and Communications, Institute for Quantum Sensing and Information Processing, Shanghai Jiao Tong University, Shanghai 200240, People's Republic of China}
\author{A. Yang}
 \affiliation {Shanghai Institute of Optics and Fine Mechanics (SIOM), Chinese Academy of Sciences, Shanghai, China}
\author{J. Han}
 \email{hanjize@cmss.chinamobile.com}
 \affiliation {China Mobile (Suzhou) Software Technology Co., Ltd., Suzhou 215163, China}
\author{Y. Wu}
 \affiliation {The 32nd Research Institute of China Electronics Technology Corporation, Shanghai, China}
\author{X. Lv}
 \email{xudonglv@siom.ac.cn}
 \affiliation {Shanghai Institute of Optics and Fine Mechanics (SIOM), Chinese Academy of Sciences, Shanghai, China}

\date{\today}

\begin{abstract}
The latency of quantum measurement and classical decoding constrains the execution of adaptive fault-tolerant circuits. In neutral-atom processors, this requirement couples spatially resolved fluorescence detection to classical inference. Here we implement cascaded readout and decoding in a heterogeneous architecture that overlaps acquisition and processing at two spatial scales. Field-programmable gate arrays process camera rows as they arrive, while graphics processors decode an acquired logical block during acquisition of the next. This heterogeneous architecture offers flexibility in the choice of error-correcting codes and decoding algorithms. Direct communication preserves \textit{microsecond}-scale transfer times for kilobyte payloads and supports interchangeable decoding algorithms. Hardware characterization and decoding benchmarks using circuit-level simulated syndromes yield a \textit{submillisecond} post-exposure processing latency for selected small-code workloads ($d = 3, 5, 7$). 
By enabling logical blocks to be decoded before full-frame readout is complete, the architecture implements low-latency syndrome decoding, realizing \textit{real-time} quantum information processing as required for fault-tolerant computation in atom arrays.

\end{abstract}

\maketitle

\section{Introduction}
\label{sec:intro}
Quantum error correction enables reliable execution of deep quantum circuits by protecting logical information against physical errors~\cite{shor1995,gottesman1998,kitaev2003}. Postselection that rejects every run with a detected error can incur exponential sampling overhead as circuits grow~\cite{andersen2020repeated,smith2024mitigating,english2025thresholds}. Real-time decoding recovers logical information from correctable runs in time for adaptive operations. It requires both a sustained processing rate that keeps pace with syndrome generation and a response time compatible with dependent logical operations~\cite{terhal2015,battistel2023realtime,fowler2012}. These requirements are distinct: corrections can often be tracked in a Pauli frame, while a later adaptive operation imposes a deadline. Real-time QEC and decoding experiments have established this principle in trapped-ion and superconducting processors~\cite{ryan2021,krinner2022,sivak2023}. Extending it to large atomic arrays requires the physical readout and classical processing to be organized together.

Neutral-atom arrays combine spatially resolved control with reconfigurable logical operations~\cite{bluvstein2022,bluvstein2024}. However, their measurement interface presents a particular timing problem. Fluorescence must be collected, digitized, and converted into site-resolved results before a decoder can infer the logical state~\cite{graham2022,madjarov2020}. Waiting for a complete camera frame makes one logical block depend on the acquisition of other blocks whose measurements may be irrelevant to its next operation. Faster inference alone cannot remove this delay. Low-latency GPU-controller interfaces, including NVQLink~\cite{nvidia_gpudirect,nvidia_cudaq_realtime}, address classical transport; camera-based atomic processors additionally require the availability of syndrome data to be coordinated with optical acquisition. The central question is how to make the logical information available as soon as its required measurements have been acquired.

Field-programmable gate array (FPGA)-based decoders achieve low latency through dedicated parallel logic, but the complexity of their implementation is constrained by on-chip resources~\cite{liyanage2024fpga,bascones2025exploring}. Graphics processing unit (GPU)-accelerated decoding offers parallel computation and software flexibility, facilitating changes in error-correcting codes and decoding algorithms~\cite{gicev2023scalable,caldwell2025platform}. These complementary strengths motivate a heterogeneous architecture that assigns streaming readout processing to an FPGA and decoding to a GPU. Realizing this advantage requires a communication interface that preserves the latency gains of both processors~\cite{caldwell2025platform}. For neutral-atom arrays, the integration must also encompass optical acquisition: measurement data should reach the decoder as each logical block becomes available, without waiting for the complete camera frame.

Here, we present and implement a heterogeneous architecture for cascaded readout and decoding. An FPGA processes the camera stream and assembles measurement records, while a GPU runs software-defined decoders for different codes~\cite{fu2021,das2020,barber2025real}. A direct peer-to-peer connection transfers these records to GPU memory without staging the data in host memory~\cite{nvidia_gpudirect}. The architecture overlaps acquisition and processing at two levels: the FPGA processes camera rows as they arrive, while the GPU decodes each acquired logical block during readout of the next. This allows decoding to begin without waiting for unrelated regions of the camera frame. 

We evaluate three representative code–decoder pairings: lookup-table (LUT) decoding of repetition codes~\cite{das2022lilliput}, belief propagation with ordered-statistics post-processing (BP-OSD) for color codes~\cite{roffe2020,panteleev2021}, and recurrent-transformer decoding of surface codes~\cite{bausch2024learning,senior2025scalable}. These pairings provide contrasting computational workloads, from table lookup and syndrome-dependent iterative decoding to neural inference with a fixed computational graph for a given input size. For each pairing, we characterize the logical error rate as a function of physical error rate and code distance, and the decoding time as a function of distance, record length, and physical error rate.


We characterize the optical readout and interconnect hardware and benchmark repetition-, surface-, and color-code decoders using circuit-level simulated syndromes~\cite{gidney2021stim}. The direct link transfers packets up to 1~KB with a round-trip transfer latency of approximately 4\,$\mu$s or less. This corresponds to an approximately 32-fold increase in payload relative to the 32-byte packet used in the initial NVQLink latency demonstration, on a comparable microsecond timescale~\cite{nvidia_gpudirect}. The component measurements yield a post-exposure processing budget below 1\,ms for selected small-code workloads. A recurrent-transformer decoder also reduces the logical error rate relative to the tested uncorrelated matching baseline~\cite{higgott2025sparse}. Together, these results identify an operating regime in which optical readout can be coupled to software-defined decoding within a millisecond-scale processing budget. A comparison with representative neutral-atom QEC experiments and recent low-latency decoding systems is provided in Table~\ref{tab:comparison} (Appendix~\ref{app:comparison}), highlighting the distinction between decoder-only benchmarks and latency measurements that include the physical readout pipeline.

The code geometry also shapes the camera-readout contribution to latency. At fixed imaging geometry, a distance-$d$ surface-code block occupies $O(d)$ image rows but contains $O(d^2)$ measured sites. With a fixed camera row time and FPGA processing that keeps pace with the row stream, this readout contribution grows linearly with $d$. Processing each logical block as its measurement data become available avoids additional waiting for unrelated image regions. Delayed logical decisions can extend circuit execution, leaving encoded states exposed to noise and potentially requiring additional error-correction cycles.

\section{System Architecture}
\label{sec:arch}

\subsection{Universal fault-tolerant architecture for neutral atom quantum computing platform}
\label{sec:arch1}
\begin{figure}[!htbp]
\centering
\includegraphics[width=\columnwidth]{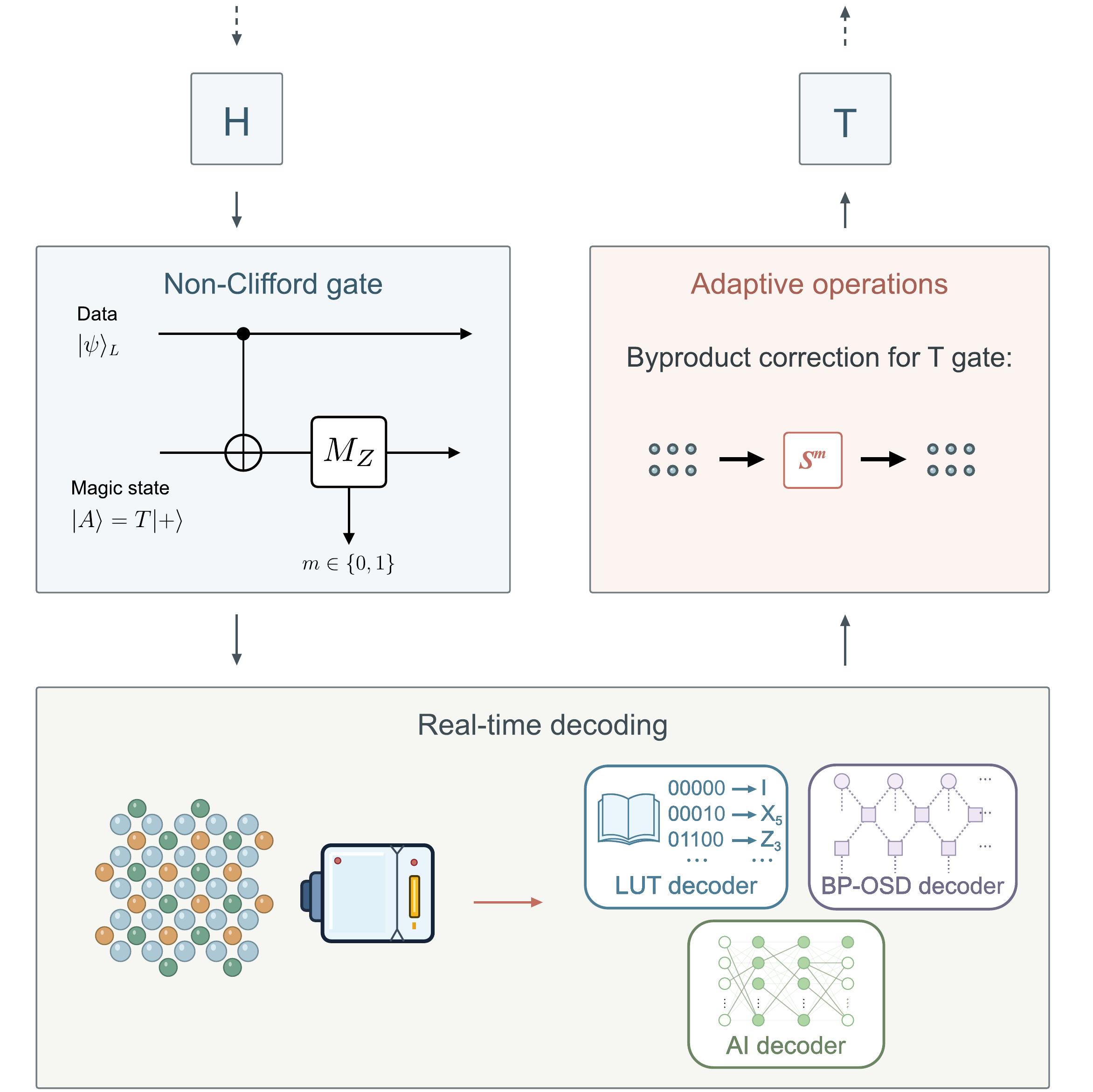}
\caption{\textbf{Schematic system architecture for real-time decoding and feedforward.} During Clifford operations, represented by the Hadamard gate H at the upper left, Pauli errors are tracked in the Pauli frame without immediate physical correction. A non-Clifford operation, illustrated by a T gate implemented through magic-state injection, requires measurement-dependent feedforward. Syndrome measurements are processed by a real-time decoder to support Pauli frame updates and the interpretation of logical measurement outcomes. The decoded information then guides the adaptive operations or gate modifications of subsequent logical operations. For the illustrated T-gate implementation, the adaptive correction takes the form $S^m$, where $m \in \{0,1\}$ is the binary feedforward decision obtained from the decoded logical measurement outcome, with the relevant Pauli-frame information taken into account. The decoded information must be available before the next non-Clifford operation whose implementation depends on it, represented by the T gate at the upper right.}
\label{fig:arch}
\end{figure}

Universal fault-tolerant quantum computation requires non-Clifford resources beyond the classically simulable stabilizer framework~\cite{gottesman1998,aaronson2004}. A standard approach is to supplement Clifford operations with T gates, which can be implemented through magic-state injection when the code does not support a transversal T gate~\cite{bravyi2012,meier2013,eastin2009}. This procedure introduces a measurement-dependent byproduct correction whose value is not known until the relevant measurement outcomes have been interpreted together with the decoded error information and the current Pauli frame. Consequently, classical decoding and feedforward must determine the necessary adaptive operations before subsequent non-Clifford gates requiring this information can proceed, making decoder latency part of the critical execution path.

Figure~\ref{fig:arch} illustrates a target system architecture that meets these decoding and feedforward requirements. In each QEC cycle, stabilizer measurement outcomes are acquired and processed by a real-time decoder, which may employ lookup-table (LUT)~\cite{das2022lilliput}, BP-OSD~\cite{panteleev2021,roffe2020}, or AI-based algorithms~\cite{bausch2024learning,senior2025scalable}. The decoder processes the syndrome to determine the most likely error and its net effect on the logical observables~\cite{fowler2012,dennis2002}. The classical control stack combines the decoder output with the relevant measurement outcomes and the existing Pauli frame to determine the frame update, byproduct correction, or gate modification. During Clifford regions, Pauli errors can be tracked entirely in software, allowing physical corrections to be deferred without interrupting circuit execution. However, for non-Clifford operations that require feedforward, the relevant decoding results must be available before the next non-Clifford gate whose implementation depends on those results can be executed; the decoding and control pipeline must therefore complete within the available execution window to avoid decoding-induced stalls. Sustained operation additionally imposes a throughput requirement: if the sustained throughput of the decoder falls below the rate of syndrome generation, the undecoded backlog grows without bound under continuous operation~\cite{terhal2015}. Such decoding and feedforward loops have been demonstrated on superconducting~\cite{sivak2023} and trapped-ion platforms~\cite{ryan2021}, and surface-code decoders operating on the order of 1\,$\mu$s are now available~\cite{senior2025scalable}. For neutral-atom architectures, where logical operations have been demonstrated~\cite{bluvstein2024}, characterizing the classical processing latency relevant to these feedforward requirements motivates the latency measurements reported below. We note that our decoder benchmarks employ memory-level QEC circuits and provide a baseline for estimating the decoding contribution to feedforward latency under comparable decoding workloads; extending this assessment to a complete magic-state injection protocol further requires accounting for its circuit-specific error propagation, logical measurements, and conditional control.

\subsection{Decoding pipeline architecture}
\label{sec:arch2}

\begin{figure*}[!htbp]
\centering
\includegraphics[width=\textwidth]{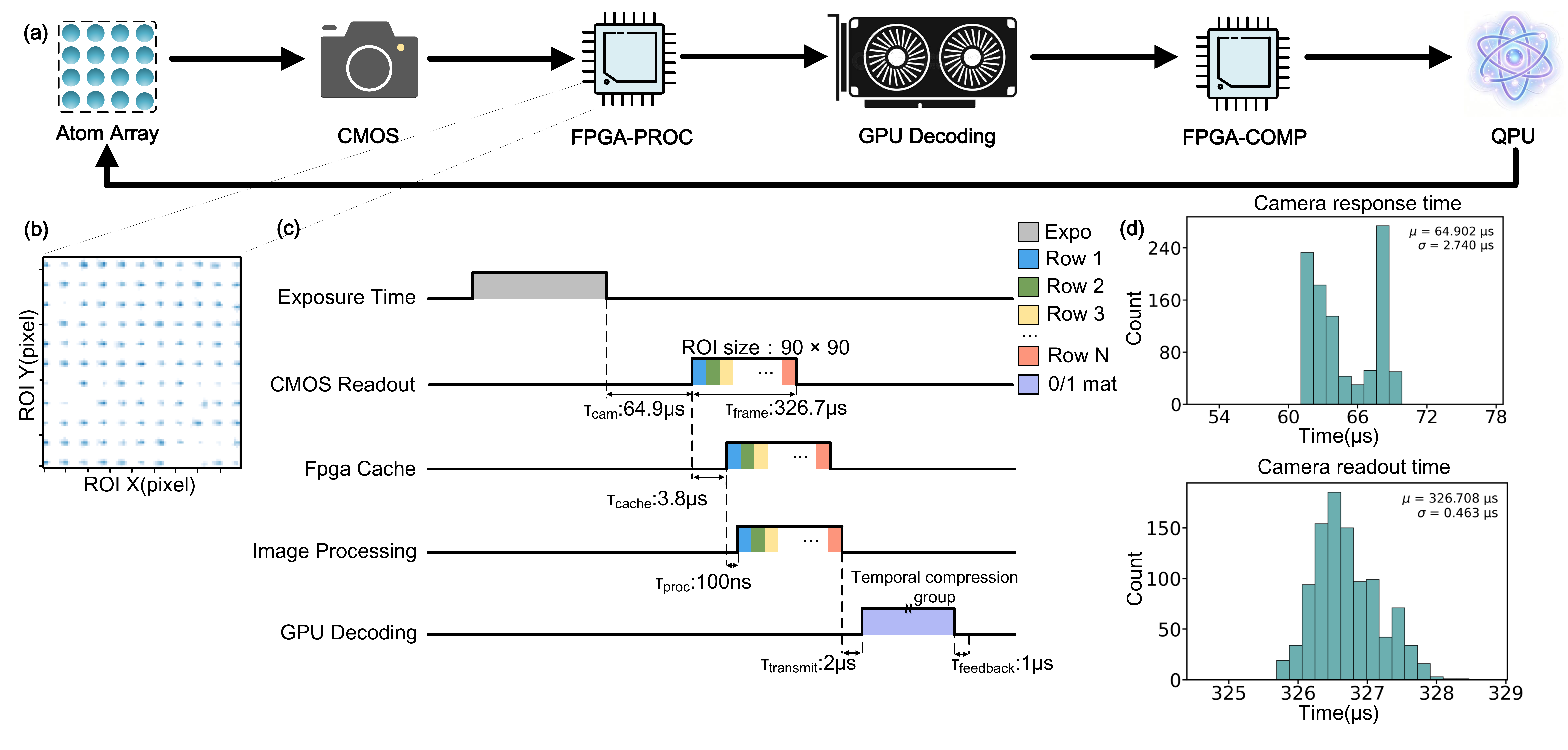}
\caption{\textbf{Real-time QEC feedback pipeline in a neutral-atom processor.} (a) Schematic of the target closed-loop path from physical readout to control actuation. The stages from fluorescence imaging through FPGA-based image processing and GPU decoding to correction feedforward are implemented in hardware in this work, whereas the final actuation onto the quantum processor is reserved for future closed-loop operation. (b) Atomic fluorescence within the camera ROI, as demonstrated in the experiment; each site yields a site-resolved ``occupied'' or ``empty'' signal. (c) Measured latency breakdown for a $90 \times 90$ pixel camera ROI, from the end of the fluorescence exposure through row-by-row CMOS readout, image preprocessing and transfer, FPGA buffering, atom reconstruction, and syndrome transfer to the GPU, to the availability of the decoded syndrome at the decoder output. (d) Histograms of the camera response time and readout time over repeated trials; $\mu$ is the sample mean and $\sigma$ is the standard deviation.}
\label{fig:pipeline}
\end{figure*}

The viability of this architecture hinges on the end-to-end speed with which the decoder returns the correction operation. Since non-Clifford gates require the execution of the feedforward stall circuit until the relevant decoding results become available, the full decoding-pipeline latency introduces additional waiting time and increases the overall circuit runtime~\cite{bravyi2005,gidney2021}. Figure~\ref{fig:pipeline}(a) sketches the complete readout-to-decode-to-actuate path for a neutral-atom quantum processor. Measurement begins with state-selective fluorescence imaging of the atoms: a CMOS camera captures a full frame in which each trap site yields a site-resolved ``occupied'' or ``empty'' signal~\cite{saffman2010,madjarov2020}. These data are first interpreted by an image-processing FPGA (FPGA-PROC), then packaged and sent to a GPU for decoding. The decoded syndrome is further forwarded to a compiling FPGA (FPGA-COMP), which drives the QPU to perform real-time feedforward control.
A single-shot fluorescence image of the atoms is captured, from which the state occupancy of the atoms within the Region of Interest (ROI) is extracted through image processing (see Figure~\ref{fig:pipeline}(b)). The stages of the pipeline and their measured latency contributions are shown in Figure~\ref{fig:pipeline}(c). Owing to architectural constraints, CMOS sensors cannot address individual pixels independently; readout proceeds row by row. To minimize redundant data transfer, the system takes advantage of the spatial partitioning of the atom array and operates in ROI mode, reading only the rows and columns that contain the target atom sites~\cite{chiu2025continuous}. A further improvement along this direction is to replace frame-based CMOS imaging with site-resolved single-photon detector arrays, where continuous photon counting combined with real-time adaptive protection eliminates the common exposure and row-sequential readout latency altogether~\cite{zeng2026fast}. We consider a distance-7 rotated surface code in the zoned architecture~\cite{bluvstein2024}, in which ancilla atoms are shuttled to a dedicated readout zone for syndrome extraction while the data atoms remain outside; for our trap spacing and imaging magnification, a $90 \times 90$ pixel ROI accommodates the 48 ancilla atoms of one logical block with margin for alignment and positional fluctuations.

Following the exposure period (Expo), the entire ROI image is read from the CMOS chip in 326.7\,$\mu$s; image preprocessing and transfer add another 64.9\,$\mu$s. FPGA buffering contributes only 3.8\,$\mu$s, and image analysis is completed in roughly 100\,ns, resulting in the occupancy matrix 0/1. The occupancy pattern is then passed to the GPU over a GPUDirect RDMA zero-copy link, reaching the GPU memory in 1.8\,$\mu$s without an intermediate copy through host memory; the decoded correction is subsequently returned to the FPGA along the same interconnect in 1.0\,$\mu$s, so that the correction instruction is available in the control FPGA with negligible overhead. A complete description of the architecture, the zero-copy data path, and the detailed latency characterization are given in Appendix~\ref{app:fpga}. In this demonstration, the decoded correction is delivered to the classical control stack but not yet actuated onto the QPU, and the closed-loop path is reserved for a future stage; we therefore do not report a readout-to-actuation latency. Instead, we characterize processing latency, defined as the time from the end of fluorescence exposure to the availability of the decoded result at the compiling FPGA input. For a $d=7$ rotated surface code decoded with the AI decoder, this latency averages 0.76\,ms.

We compare the measured latency with the relevant QPU operation timescales. State-of-the-art neutral-atom systems execute two-qubit Rydberg gates in $\sim$0.5--1\,$\mu$s and single-qubit gates on the order of 100\,ns, while coherent transport of atoms takes several hundred microseconds~\cite{bluvstein2024,evered2023,bluvstein2022}; the relevant quantum processor clock speed is on the order of milliseconds per round~\cite{bluvstein2026faulttolerant}. Our measured latency falls within this millisecond-scale cycle time, supporting the feasibility of real-time decoding. Operating the pipeline at a latency comparable to the quantum processor clock provides a foundation for integrating non-Clifford gates into a fault-tolerant workflow~\cite{bravyi2005,litinski2019}.

\section{Decoding Performance: Accuracy and Latency}
\label{sec:perf}

We evaluate the decoding accuracy and latency of three representative code--decoder pairings: BP-OSD decoding for color codes, a recurrent-transformer decoder (hereafter referred to as the AI decoder) for surface codes, and LUT decoding for repetition codes as a low-latency reference (Appendix~\ref{sec:perf_lut}). These algorithms exhibit distinct computational characteristics, ranging from near-constant-time lookup and syndrome-dependent iterative decoding to neural-network inference with a fixed computational graph, allowing us to assess their performance within a common FPGA--GPU processing architecture.

For all three code families, syndrome data are generated using Stim~\cite{gidney2021stim} under circuit-level depolarizing noise and injected into the FPGA--GPU data path shown in Fig.~\ref{fig:pipeline}. We characterize the logical error rate (LER) as a function of the physical error rate (PER) and the code distance $d$, and the GPU decoding time as a function of $d$, the number of stabilizer-measurement rounds $r$ and PER. LER per round, $p_L$, is extracted from the decay of logical fidelity over repeated rounds of syndrome extraction (Appendix~\ref{app:sre}). Motivated by the millisecond-scale syndrome-extraction cycles of current neutral-atom processors~\cite{bluvstein2024,zhou2025resource}, we adopt a 1\,ms reference budget for GPU decoding, distinct from the end-to-end processing latency that additionally includes camera readout, FPGA processing, and data transfer.

\subsection{BP-OSD Decoding of Color Codes}
\label{sec:perf_bposd}

The 2D color code supports transversal implementation of the complete Clifford gate set~\cite{bombin2006,landahl2011,kubica2015} (Appendix~\ref{app:codes}). However, its parity-check matrices contain numerous short cycles, and the code exhibits substantial degeneracy, both of which can challenge conventional belief-propagation decoding. We therefore employ BP-OSD, which combines belief propagation with ordered-statistics post-processing~\cite{roffe2020,panteleev2021}, and implement it on the GPU using CUDA for $d=3,5,7$.

\begin{figure}[!htbp]
\centering
\includegraphics[width=\columnwidth]{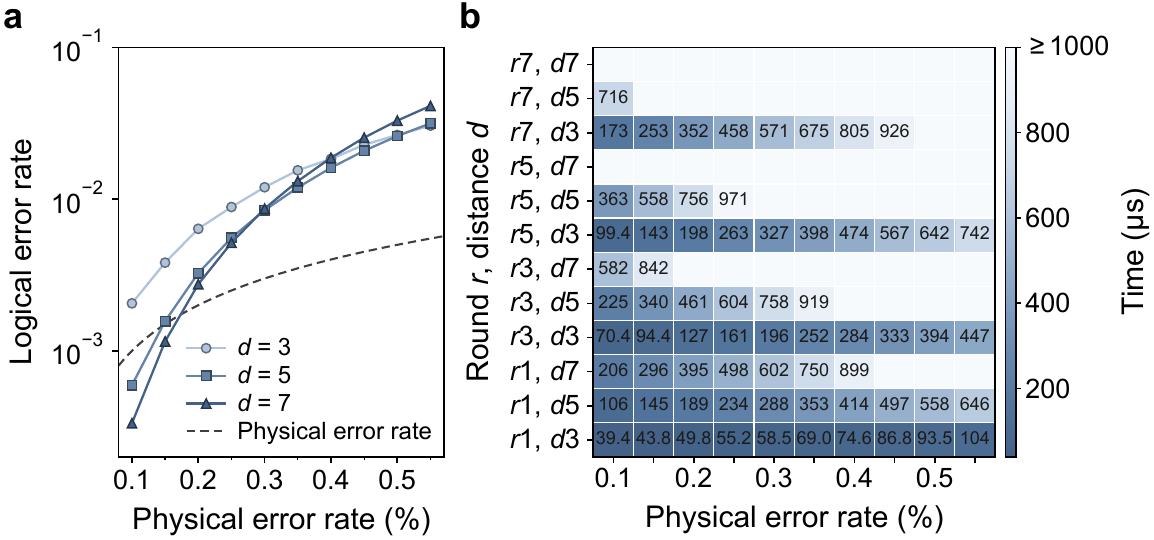}
\caption{\textbf{Performance of the BP-OSD decoder for the color code.} (a) Logical error rate (per round) for Z-basis decoding as a function of physical error rate. Each value is obtained by fitting the decay of the logical fidelity over $r = 1$ to $15$ syndrome-extraction rounds. (b) Average GPU decoding time per shot as a function of physical error rate for different combinations of code distance $d$ and syndrome-extraction rounds $r$. Blank cells indicate configurations for which the average decoding time exceeds the 1\,ms reference budget. For each parameter setting, logical fidelities are estimated and decoding times are averaged over $10^5$ shots.}
\label{fig:color}
\end{figure}

Figure~\ref{fig:color}(a) shows the resulting LER as a function of PER. Increasing the code distance suppresses the LER in the low-noise regime: the $d=5$ and $d=7$ codes reach a logical error rate per round below the physical error rate for $p \leq 0.15\%$ and $p \leq 0.17\%$, respectively, whereas the $d=3$ code does not satisfy $p_L < p$ anywhere in the investigated range. The separation between distances narrows as $p$ increases, and the $d=5$ and $d=7$ curves cross at $p \sim 0.30\%$. The crossover suggests that the benefit of increasing code distance becomes less pronounced at higher physical error rates under the present decoding configuration. However, it should not be interpreted as a threshold estimate, since finite-size effects and decoder-specific limitations may also contribute.

The GPU decoding time, plotted in Fig.~\ref{fig:color}(b), generally increases with code distance $d$, the number of measurement rounds $r$, and PER. Increasing $d$ or $r$ enlarges the decoding problem and raises the computational cost of both BP and OSD. In contrast, increasing PER leaves the dimensions of the decoding problem unchanged but alters the syndrome distribution, leading to input-dependent variations in decoding time.

While small-code configurations remain within the 1\,ms reference budget on average, larger distances and longer measurement records approach or exceed this limit. Moreover, the input-dependent runtime of BP-OSD introduces an additional consideration for real-time decoding: meeting a fixed feedback deadline requires controlling not only the average decoding time but also its variability. These results motivate further optimization of BP-OSD to reduce decoding latency and improve its suitability for real-time operation at larger code distances.

\subsection{Hybrid Recurrent-Transformer AI Decoding of Surface Codes}
\label{sec:perf_ai}

For the rotated surface code (Appendix~\ref{app:codes}), we adopt the recurrent-transformer decoder architecture of Ref.~\cite{senior2025scalable}, hereafter referred to as the AI decoder. The network learns spatiotemporal error correlations from detection-event sequences through offline training. We use the same network architecture at all three distances and transfer trained models from smaller to larger codes (Appendix~\ref{app:ai_decoder}). We benchmark it against minimum-weight perfect matching (MWPM)~\cite{higgott2025sparse,dennis2002}, a widely used reference decoder for surface codes.

\begin{figure}[!htbp]
\centering
\includegraphics[width=\columnwidth]{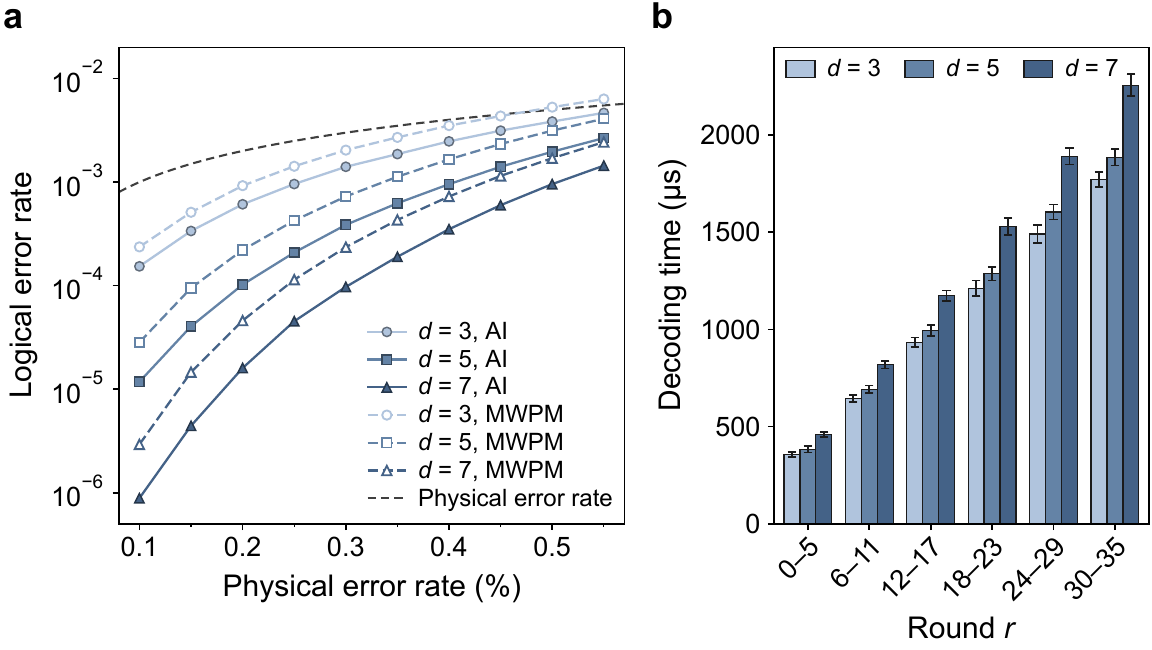}
\caption{\textbf{Performance of the AI decoder for the rotated surface code.} (a) Logical error rate (per round) for Z-basis decoding versus physical error rate for the AI decoder (filled symbols, solid lines) and MWPM decoder (open symbols, dashed lines). Each logical error rate per round is obtained by fitting the decay of the logical fidelity over $r = 1$ to $30$ syndrome-extraction rounds. The number of shots used to evaluate the logical fidelity ranges from \(10^5\) to \(10^7\) per parameter setting. (b) Total GPU decoding time of the AI decoder versus the number of syndrome-extraction rounds $r$ at a physical error rate of $0.5\%$. The complete detection-event record is processed in a single forward pass, so the reported time corresponds to decoding an entire record. The decoder compresses \(k=6\) consecutive rounds into one group, with records padded to a multiple of \(k\). The decoding time is therefore approximately constant within each labeled range of \(r\) (e.g., 0–5, 6–11; both endpoints included) and increases between successive ranges. Each bar represents the mean GPU decoding time over $10^5$ inference trials at a representative round count ($r=1,6,12,18,24,30$), with error bars indicating the standard deviation across trials. 
}
\label{fig:ai}
\end{figure}

Figure~\ref{fig:ai}(a) compares the LER per round of the two decoders, extracted by fitting the decay of the logical fidelity over $r = 1$ to $30$ syndrome-extraction rounds. The AI decoder consistently achieves a lower LER than the tested MWPM baseline across all investigated code distances and physical error rates, with a more pronounced advantage at larger distances. Both decoders exhibit improved logical-error suppression with increasing code distance over the investigated noise range. The improved performance is consistent with the AI decoder's ability to learn spatiotemporal error correlations, although the MWPM baseline used here does not incorporate correlated reweighting.

To assess the decoding latency, Fig.~\ref{fig:ai}(b) plots the total GPU decoding time against the number of syndrome-extraction rounds $r$, with the complete detection-event record processed in a single inference call. Because the decoder compresses groups of $k=6$ consecutive rounds into a single representation~\cite{senior2025scalable}, the decoding time increases in discrete steps at group boundaries (Appendix~\ref{app:ai_decoder}). Each additional group of six rounds increases the decoding time by approximately $290$, $310$, and $360\,\mu$s for $d=3$, $5$, and $7$, respectively. Despite a sixfold increase in the number of stabilizers from $d=3$ to $d=7$, the incremental decoding time increases by only about $24\%$. These additional processing costs remain below the 1\,ms reference decoding budget, suggesting computational headroom for processing longer syndrome records.

\section{Discussion and Conclusion}
\label{sec:concl}

In this work, we presented a heterogeneous FPGA–GPU architecture that integrates image readout, FPGA image processing, and GPU decoding for neutral-atom quantum error correction. Table~\ref{tab:latency} gives the stage-by-stage latency budget and the code distance dependence for the code–decoder pairings considered here. Hardware characterization combined with simulated syndrome benchmarks identifies camera readout and GPU decoding as the largest contributions to the open-loop post-exposure latency. By resolving the interval from the end of exposure to the return of a decoded result, this characterization provides a basis for coordinating optical acquisition, classical processing, and subsequent adaptive operations.


\begin{table}[!htbp]
\caption{\textbf{\boldmath Measured latencies and their scaling with the code distance $d$ for the codes considered.} Decoding times and their scaling with $d$ are reported for the decoder–code combinations used in this work: LUT decoding for the repetition code, BP-OSD decoding for the color code, and AI decoding for the rotated surface code. All other latency values refer to the rotated surface code at $d=7$, and their scaling with $d$ applies to this code family.}
\label{tab:latency}
\footnotesize
\setlength{\tabcolsep}{3pt}
\renewcommand{\arraystretch}{1.25}
\begin{ruledtabular}
\begin{tabular}{@{}>{\raggedright\arraybackslash}p{2.65cm}>{\raggedright\arraybackslash}p{2.35cm}>{\raggedright\arraybackslash}p{2.55cm}@{}}
Stage & Latency & Scaling with $d$ \\
\colrule
CMOS ROI readout & 326.7\,$\mu$s & $\propto d$ (row-sequential) \\
Preprocessing + transfer & 64.9\,$\mu$s & $\propto d^2$ (pixel count) \\
FPGA buffering & 3.8\,$\mu$s & $\propto d^2$ \\
Occupancy analysis & 100\,ns &  constant \\
FPGA$\rightarrow$GPU transmit & 1.8\,$\mu$s & $\propto d^2$ payload \\
GPU$\rightarrow$FPGA feedback & 1.0\,$\mu$s & constant \\
LUT decoding & 82\,$\mu$s & constant\\
BP-OSD decoding & grows with $d$, $r$ & superlinear ($\gtrsim d^2$) \\
AI decoding & 360\,$\mu$s / 6 rounds ($d{=}7$)& sublinear ($+24\%$, $d{=}3\to7$) \\
\end{tabular}
\end{ruledtabular}
\end{table}

For a given logical block, the post-exposure latency \(L_{\mathrm{post}}\)
is the time from the end of the final fluorescence exposure to the availability
of the decoded result at the input of FPGA-COMP. For the complete-record decoding considered here,
the timing budget is
\begin{equation}
\label{eq:qec-post-exposure-latency}
L_{\mathrm{post}}
=
L_{\mathrm{rp}}
+ L_{\mathrm{FPGA}\to\mathrm{GPU}}
+ L_{\mathrm{dec}}
+ L_{\mathrm{GPU}\to\mathrm{FC}}
\end{equation}
The first term is the time from the end of exposure until the decoder input
data for that block are ready for transfer (\(L_{\mathrm{rp}}\)). This
interval includes the camera response, ROI readout, and FPGA processing,
with concurrent row acquisition and image processing counted once. The
remaining terms are the time to transfer the input data from the
image-processing FPGA to the GPU (\(L_{\mathrm{FPGA}\to\mathrm{GPU}}\)),
the time from GPU input availability to decoder output readiness
(\(L_{\mathrm{dec}}\)), and the time to return the decoded result from the
GPU to FPGA-COMP (\(L_{\mathrm{GPU}\to\mathrm{FC}}\)).
In Table~I, \(L_{\mathrm{rp}}\) and \(L_{\mathrm{dec}}\)
dominate the budget in Eq.~\eqref{eq:qec-post-exposure-latency}, with
camera ROI readout accounting for most of \(L_{\mathrm{rp}}\).

The camera readout contribution to $L_{\mathrm{rp}}$ is set by the number of ROI rows and the per-row readout time. At fixed trap spacing and magnification, the row count for a surface-code block grows linearly with the code distance $d$, and so does the readout contribution, even when the FPGA processes each row as it arrives. The same geometry fixes the FPGA-to-GPU transfer payload through the ROI dimensions, while the decoder input size sets the decoding cost. Matching the ROI to the required ancilla sites minimizes $L_{\mathrm{rp}}$; as a further improvement, site-resolved photon-counting arrays would remove row-sequential readout altogether~\cite{zeng2026fast}.

The dependence of $L_{\mathrm{dec}}$ on the code distance $d$ differs between the code-decoder pairs tested. LUT decoding of repetition codes takes approximately the same time across the tested distances. BP-OSD decoding of color codes takes longer as code distance, record length, or physical error rate increases, making both mean latency and syndrome-dependent variability relevant to feedback deadlines. For the surface-code with AI decoder, the additional time needed to process each extra group of rounds increases only modestly with $d$ over the tested range.

During continuous circuit execution, the stages need not all contribute sequentially to the interval between successive logical operations. The cascaded architecture overlaps row acquisition with FPGA image processing and allows GPU decoding of an acquired logical block while the camera reads the next block. Across successive QEC rounds, decoding of earlier data could also overlap later syndrome extraction or independent atom transport, with syndrome history maintained across processing windows~\cite{zhou2025resource}. Clifford operations can often proceed while corrections are tracked in the Pauli frame, whereas an adaptive operation requires the decoded information on which its implementation depends~\cite{chamberland2018fault}. This distinction allows processing to be scheduled around the operations that impose a feedforward deadline, hiding part of the classical latency while maintaining a processing rate that keeps pace with syndrome generation. Although the present decoder timings use complete records supplied at once, the architecture motivates temporal streaming and shared-GPU scheduling across logical qubits as routes toward sustained QEC in larger processors~\cite{skoric2023parallel,maurya2026case}.

The significance of the measured latency at the system-level depends on its relation to the measurement and transport timescales of the processor. Fluorescence readout requires an exposure interval to collect sufficient photons for state discrimination. For example, mid-circuit fluorescence imaging with a $500\,\mu\mathrm{s}$ exposure has been demonstrated in neutral-atom arrays~\cite{bluvstein2024}. Transport time also enters the timing budget in the commonly adopted zoned architecture, where atoms are shuttled between storage, entangling, and readout zones during circuit execution. In a $d=5$ surface-code experiment, moving the ancilla qubits to the storage zone and bringing the next group to the entangling zone took a total of $2.57\,\mathrm{ms}$ per round~\cite{bluvstein2026faulttolerant}. In a fast-readout implementation, classical processing can constitute a substantial part of the interval before a decoded result becomes available. Its effect on the experimental execution time is determined by how much of this processing remains on the causal critical path. As discussed above, independent transport and quantum operations can provide a window for classical  processing before the next operation that depends on the decoded result. Reducing the post-exposure latency can therefore help processing fit within this window or shorten the remaining feedforward stall.

The decoder benchmarks characterize $L_{\mathrm{dec}}$ for complete
syndrome records. During continuous circuit execution, the stages that make
up $L_{\mathrm{post}}$ in Eq.~\eqref{eq:qec-post-exposure-latency} can
overlap with work on other blocks. For example, the GPU can decode one
block while the camera reads the next. Across QEC rounds, a streaming
decoder could also process earlier data during later syndrome extraction
or independent atom transport, while maintaining syndrome history across
processing windows~\cite{zhou2025resource,skoric2023parallel,maurya2026case}. Clifford operations can often proceed
with Pauli-frame tracking, while a conditional operation must wait for
the decoded information on which it depends~\cite{chamberland2018fault}.

For the closed-loop timing model, we take the end of the final required
exposure as the time origin. Let $W_{\mathrm{QPU}}$ be the earliest time at
which the quantum processor could start the conditional operation if the
control instruction were already available. This time is set by the completion of the required transport and preceding quantum operations. Let
$L_{\mathrm{cmd}}$ be the time from receipt of the decoded result at the FPGA-COMP until the control instruction is ready. This step may include interpreting a logical measurement outcome in the current Pauli frame and selecting the corresponding correction.

When command preparation and quantum-side preparation can proceed
independently, the modeled time from the start of the final exposure to the
onset of the conditional operation is
\begin{equation}
T_{\mathrm{loop}}^{\mathrm{model}}
=
T_{\mathrm{expo}}
+ \max\!\left\{L_{\mathrm{post}}+L_{\mathrm{cmd}},\,W_{\mathrm{QPU}}\right\}
+ L_{\mathrm{launch}}
\end{equation}
where $T_{\mathrm{expo}}$ is the duration of the final exposure and
$L_{\mathrm{launch}}$ is the delay from readiness on both sides to the
onset of physical actuation. A $500\,\mu\mathrm{s}$ exposure has been
demonstrated for mid-circuit fluorescence imaging~\cite{bluvstein2024}, providing a
reference scale for $T_{\mathrm{expo}}$.

In a zoned architecture, atoms are shuttled between storage, entangling,
and readout zones. Transport prior to exposure contributes to the duration of the QEC cycle
. After exposure, transport required for the conditional operation
contributes to $W_{\mathrm{QPU}}$ if it can proceed independently of the
decoded result. Ref.~\cite{bluvstein2026faulttolerant} reports $2.57\,\mathrm{ms}$ per
round for moving one ancilla group to storage and bringing the next into
the entangling zone. The portion available for overlap depends on the
schedule.

If the required atom transport and preceding quantum operations finish before the control instruction is ready, the quantum processor must wait for that instruction. The resulting feedforward stall is
\begin{equation}
T_{\mathrm{stall}}
=
\max\!\left\{0,\,L_{\mathrm{post}}+L_{\mathrm{cmd}}-W_{\mathrm{QPU}}\right\}
\end{equation}
The experiments reported here characterize $L_{\mathrm{post}}$. The
closed-loop model determines when reducing this latency shortens the
circuit schedule by reducing the feedforward stall. This is particularly relevant to fast transversal architectures, in which reaction time can limit algorithm execution despite short quantum-operation sequences~\cite{zhou2025resource}. As readout and transport become faster, coordinating these stages will become increasingly important for
translating faster physical operations into faster logical computation.

The heterogeneous FPGA--GPU architecture and the corresponding latency characterization represent a step toward fully integrated real-time QEC on neutral-atom platforms. As neutral-atom processors scale and implement deeper fault-tolerant circuits, both sustained decoding throughput and timely availability of logical decisions will become
increasingly important. The next steps are to integrate the pipeline with experimental stabilizer measurements, evaluate larger codes and circuit-specific decoding workloads, and demonstrate closed-loop feedforward. These developments will establish how the measured processing performance translates into the execution time and logical performance of fault-tolerant neutral-atom computation.
\vspace{0.8\baselineskip}

\begin{acknowledgments}
\begin{CJK*}{UTF8}{gbsn}
This work was supported by National Key R\&D Program of China (Grant No. 2025YFF0515500, 2025YFF0515502 and 2025YFF0515504),
Strategic Frontier Project of STCSM (Grant No. 25DP2600100). J. H. acknowledges the Natural Science Foundation of Jiangsu Province (No. BK20250404), and the Youth Science and Technology Talent Support Project of Jiangsu Province (No. JSTJ-2025-600). We also thank Joonhee Choi for helpful discussions. The authors acknowledge the use of Codex and ChatGPT 6.1 Sol to assist with preparing Figure~\ref{fig:arch} and improving the readability of the manuscript. All AI-assisted content was reviewed and verified by the authors.
\end{CJK*}
\end{acknowledgments}

\bibliography{main}

\clearpage

\appendix

\section{FPGA--GPU Decoding Architecture and Latency Characterization}
\label{app:fpga}

\begin{figure}[!htbp]
\centering
\includegraphics[width=\columnwidth]{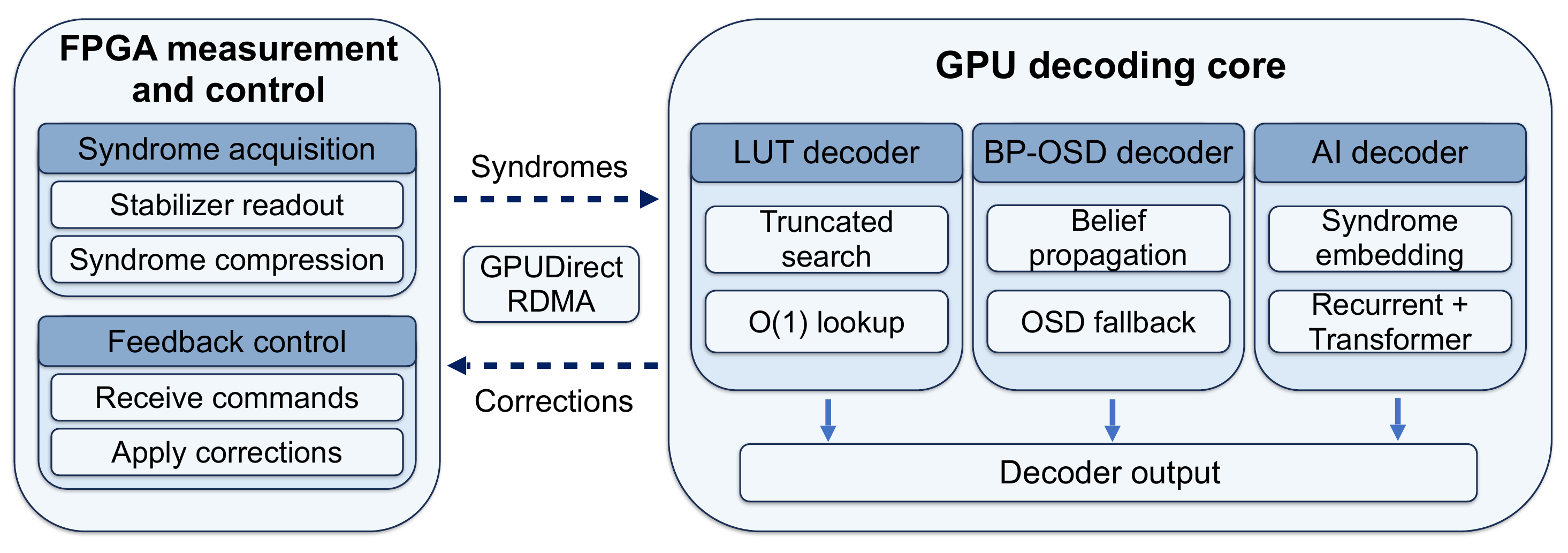}
\caption{\textbf{FPGA–GPU architecture for modular real-time decoding.} The FPGA manages stabilizer measurement, syndrome compression, and waveform control. Decoding is offloaded to a GPU through a GPUDirect RDMA zero-copy interface, where interchangeable LUT, BP-OSD, and AI decoders operate on the incoming syndrome stream. Independent of the selected decoding algorithm, the GPU returns a unified correction instruction to the FPGA, enabling adaptive feedforward during fault-tolerant quantum computation.}
\label{fig:fpga_gpu}
\end{figure}

Rather than sacrificing algorithmic flexibility by moving inference onto fixed-function FPGA logic, the architecture in Fig.~\ref{fig:pipeline} keeps the decoding core on a GPU. This choice is well supported by recent system-level studies of real-time QEC: a GPU-hosted decoder can simultaneously support multiple decoding strategies and can be updated or swapped entirely in software without hardware refabrication \cite{das2020,barber2025real}, which is invaluable while quantum codes and noise models are still evolving \cite{fowler2012,chamberland2020}. To preserve this flexibility while still minimizing wall-clock latency, the natural optimization target is the communication path between the FPGA-based measurement subsystem and the GPU, as shown in Fig.~\ref{fig:fpga_gpu}.

Figure~\ref{fig:fpga_gpu} traces the data path of one decoding loop. The FPGA
reads out the stabilizers, compresses the resulting syndrome, and transfers it
directly into GPU memory over a GPUDirect RDMA interface. Because the
transfer is zero-copy, it avoids the intermediate buffering in host memory
required by a conventional host-staged path. The selected decoder then processes
the syndrome, and its output is returned to the FPGA as a correction instruction.
The time spent in the two directions defines the transmission latency
$T_{\scriptscriptstyle\text{transmit}}$ and the feedback latency
$T_{\scriptscriptstyle\text{feedback}}$.

\begin{figure*}[!htbp]
\centering
\includegraphics[width=0.9\textwidth]{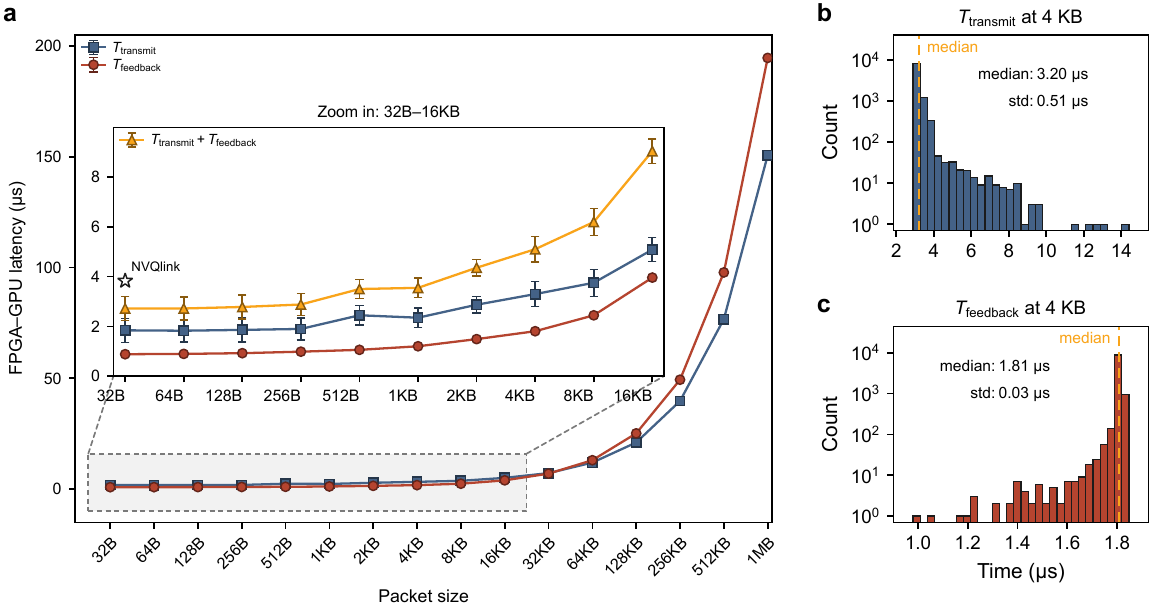}
\caption{\textbf{Measured FPGA--GPU transport latency versus packet size, with 10\,000
trials per size.} (a) One-way latencies $T_{\scriptscriptstyle\text{transmit}}$
and $T_{\scriptscriptstyle\text{feedback}}$ from 32~B to 1~MB. The inset shows
the range from 32~B to 16~KB together with the round-trip latency
$T_{\scriptscriptstyle\text{transmit}}+T_{\scriptscriptstyle\text{feedback}}$,
which has a mean of 2.72\,$\mu\mathrm{s}$ at 32~B and 3.55\,$\mu\mathrm{s}$ at
1~KB. The star marks the 3.84\,$\mu\mathrm{s}$ round-trip latency reported for
NVQLink at a 32-byte payload \cite{nvidia_gpudirect}. (b, c) Latency histograms
with logarithmic counts at a packet size of 4~KB for
(b) $T_{\scriptscriptstyle\text{transmit}}$ and
(c) $T_{\scriptscriptstyle\text{feedback}}$. Dashed lines mark the medians.}
\label{fig:latency}
\end{figure*}

Figure~\ref{fig:latency} shows both latencies for packet sizes from 32~B to
1~MB. The central result is that transport latency grows only weakly over the
payload range relevant to real-time decoding. The mean round-trip latency
$T_{\scriptscriptstyle\text{transmit}}+T_{\scriptscriptstyle\text{feedback}}$
is 2.72\,$\mu\mathrm{s}$ at 32~B and 3.55\,$\mu\mathrm{s}$ at 1~KB, so a 32-fold
increase in payload costs only 0.83\,$\mu\mathrm{s}$. Above
approximately 32~KB, latency scales nearly linearly with packet size. 

These packet sizes map directly onto decoding workloads. A distance-$d$ rotated
surface-code logical qubit produces $d^2-1$ syndrome bits per round. Neglecting
packet headers and syndrome compression, a single 32~B packet (256 bits) holds
one round for one logical qubit up to $d=15$, whereas a 1~KB packet (8192 bits)
holds one round for 18 logical qubits at $d=21$, or a full window of $d$ rounds
for one logical qubit up to $d=19$.

For comparison, at a 32-byte packet size our mean round-trip latency of
2.72\,$\mu\mathrm{s}$ is below the 3.84\,$\mu\mathrm{s}$ reported for NVQLink
\cite{nvidia_gpudirect} (star marker in Fig.~\ref{fig:latency}). NVIDIA has since
reported a transport-only median of 2.92\,$\mu\mathrm{s}$, with the payload size
not stated \cite{nvidia_cudaq_realtime}. These comparisons are indicative only,
as our link is a tightly coupled PCIe interconnect whereas NVQLink uses an
Ethernet fabric, and the measurement methods differ.

\section{Positioning of this work among decoding systems for neutral-atom QEC}
\label{app:comparison}

\begin{table*}[!htbp]
\caption{\textbf{Comparison of neutral-atom QEC experiments and decoders compatible with millisecond-scale cycle times.} The upper group lists neutral-atom QEC experiments with offline decoding; the middle group lists decoders benchmarked primarily on simulated data. ``Readout in timing'' indicates whether the reported timing includes the physical measurement and readout chain. ``Throughput'' gives decoding time normalized as reported in each work, where a round (equivalently, a cycle) is one round of syndrome extraction and a shot is one complete experiment comprising all rounds. ``Latency'' is the time interval from the final measurement to the decoder output. Timings are quoted as reported and were measured on different hardware (TPU for AQ2, GPU for Cascade and the transformer decoder, single-core CPU for beam search); they are not directly comparable (e.g., batched vs.\ single-shot, mean vs.\ worst-case). For this work, simulated syndromes are injected into the camera--FPGA--GPU readout and decoding pipeline of a Cs atom array; the throughput is the incremental decoding time per additional six-round group with all rounds supplied at once, and the latency is the sum of separately measured stages from the end of the final camera exposure to the decoder output. 
Abbreviations: AQ2, AlphaQubit~2; BB, bivariate bicycle (code); CNN, convolutional neural network; $d$, code distance; exp., experimental; FPGA, field-programmable gate array; ML, machine learning; MLE, most-likely-error (decoder); MSD, magic state distillation; SC, superconducting; sim., simulated. $[[n,k,d]]$ denotes a code with $n$ physical qubits, $k$ logical qubits and distance $d$.}
\label{tab:comparison}
\setcellgapes{4pt}
\makegapedcells
\begin{ruledtabular}
\begin{tabular}{lllllll}
\makecell[lt]{Work} &
\makecell[lt]{Platform /\\ data} &
\makecell[lt]{Code} &
\makecell[lt]{Decoder} &
\makecell[lt]{Readout\\ in timing} &
\makecell[lt]{Throughput} &
\makecell[lt]{Latency} \\
\colrule
\makecell[lt]{Bluvstein \textit{et al.}~\cite{bluvstein2026faulttolerant}} &
\makecell[lt]{$^{87}$Rb,\\ exp.\ data} &
\makecell[lt]{Surface,\\ $d=3,5$} &
\makecell[lt]{MLE + ML,\\ offline} &
$\times$ &
\makecell[lt]{$\sim$0.33\,$\mu$s/shot\\ (batched)} &
--- \\
\makecell[lt]{Atom Computing~\cite{computing2026quantum}} &
\makecell[lt]{$^{171}$Yb,\\ exp.\ data} &
\makecell[lt]{Twisted\\ toric} &
\makecell[lt]{PyMatching,\\ offline} &
--- &
--- &
--- \\
\makecell[lt]{Sales Rodriguez \textit{et al.}\\ + NVIDIA~\cite{sales2025experimental,NVIDIA2025}} &
\makecell[lt]{Neutral atom,\\ exp.\ data} &
\makecell[lt]{$[[7,1,3]]$ color\\ (MSD)} &
\makecell[lt]{Transformer,\\ offline} &
$\times$ &
\makecell[lt]{$<1$~ms/shot\\ ($d=3$)} &
--- \\
\colrule
\makecell[lt]{Senior \textit{et al.} (AQ2)~\cite{senior2025scalable}} &
\makecell[lt]{Sim.\ +\\ SC exp.\ data} &
\makecell[lt]{Surface,\\ color} &
\makecell[lt]{Recurrent\\ transformer} &
$\times$ &
\makecell[lt]{$<100\,\mu$s/cycle} &
\makecell[lt]{$<500\,\mu$s} \\
\makecell[lt]{Gu \textit{et al.} (Cascade)~\cite{gu2026scalable}} &
\makecell[lt]{Sim.} &
\makecell[lt]{Surface,\\ BB} &
\makecell[lt]{CNN} &
$\times$ &
\makecell[lt]{$\sim$40\,$\mu$s/cycle} &
--- \\
\makecell[lt]{Ye \textit{et al.}~\cite{ye2026beam}} &
\makecell[lt]{Sim.} &
\makecell[lt]{$[[144,12,12]]$\\ BB} &
\makecell[lt]{Beam search} &
$\times$ &
\makecell[lt]{270\,$\mu$s/round\\ (amortized mean)} &
--- \\
\colrule
\makecell[lt]{\textbf{This work}} &
\makecell[lt]{Cs array\\ readout hardware,\\ sim.\ syndromes} &
\makecell[lt]{Surface,\\ color} &
\makecell[lt]{AI decoder} &
$\checkmark$ &
\makecell[lt]{$\sim$360\,$\mu$s\\ per 6 rounds\\ ($d=7$)} &
\makecell[lt]{$\sim$0.76\,ms\\ ($d=7$)} \\
\end{tabular}
\end{ruledtabular}
\end{table*}

To place the end-to-end latency reported in this work in context, Table~\ref{tab:comparison} summarizes neutral-atom QEC experiments with offline decoding, together with recent decoders whose reported speed is compatible with the millisecond-scale cycle times of neutral-atom processors. 
For each work, we list the decoder, whether the reported timing includes the physical readout chain, and the reported throughput and latency.

We distinguish throughput, the decoding time normalized per shot, round or cycle, from latency, the interval between the final measurement and the decoder output. Because the reported values differ in batching, normalization, hardware and model size, they are not directly comparable; for example, the $0.33\,\mu$s per shot of Ref.~\cite{bluvstein2026faulttolerant} refers to batched inference of a single ML model, whereas the $\sim$40\,$\mu$s per cycle of Cascade~\cite{gu2026scalable} is an unbatched value.
Crucially, none of the previous timings includes the physical readout chain; for instance, the sub-$500\,\mu$s latency of AQ2~\cite{senior2025scalable} covers the decoder alone.
In contrast, the $ 0.76$\,ms latency reported here ($d=7$) spans the full chain from the end of the final camera exposure to the FPGA-COMP input, executed on the FPGA–GPU readout and control hardware integrated with a Cs atom array. As simulated syndromes are injected into this pipeline, this value characterizes the system-level latency of readout and decoding rather than real-time decoding of experimental syndromes.

\section{Quantum Error-Correcting Codes}
\label{app:codes}
The three code families below, repetition, color, and (rotated) surface, share a common structure despite differing lattice geometry. Each distance-$d$ code with $N$ data qubits requires $N-1$ ancilla qubits (one per independent stabilizer), for $2N-1$ qubits total, and each corrects up to $\lfloor (d-1)/2 \rfloor$ physical errors. Stabilizer measurements in all three are performed the same way: an ancilla qubit is prepared in the appropriate basis state ($|+\rangle$ for an X-type stabilizer, $|0\rangle$ for a Z-type stabilizer), CNOT gates couple it to its associated data qubits, and a measurement of the ancilla in the corresponding basis extracts the error syndrome without collapsing the encoded state. The codes differ in lattice dimensionality, and in stabilizer weight and placement.
\subsection{Repetition code}
The quantum repetition code encodes a logical qubit across $n=d$ physical qubits by distributing quantum information into a GHZ-entangled state, where $d$ is the code distance. The bit-flip variant protects against X errors using Z-type stabilizers; an analogous phase-flip variant, related by a transversal Hadamard, protects against Z errors using X-type stabilizers. For a distance-d code there are d-1 stabilizers,
$$S_i = Z_i Z_{i+1}, \qquad i = 1, \dots, d-1.$$
The logical basis states are $$|0\rangle_L \equiv |0\rangle^{\otimes d}$$ and $$|1\rangle_L \equiv |1\rangle^{\otimes d}$$ prepared by applying controlled-X gates from the first qubit to the remaining $d-1$ qubits. The logical operators are $$X_L = X_1 X_2 \cdots X_d$$ and $$Z_L = Z_1$$ (equivalently any single $Z_i$, since they act identically on the codespace). A distance-$d$ instance requires d data qubits and $d-1$ ancilla qubits ($2d-1$ total); for $d=3$ this is 3 data and 2 ancilla qubits (5 total), and we use $d=3,5,7$ in this work.
Its simplicity makes the repetition code the standard pedagogical starting point, though its 1D connectivity and single error type per variant limit it to illustrating the syndrome-extraction principle rather than practical error correction.

\subsection{Color Code}

We consider the two-dimensional (2D) color code. The 2D color code is a topological stabilizer code defined on a trivalent, three-face-colorable lattice (typically a triangular patch of the 6.6.6 lattice) in which every face simultaneously carries one $X$-type and one $Z$-type stabilizer, both acting on the same set of data qubits. Unlike the surface code, where $X$- and $Z$-stabilizers occupy separate plaquettes, this shared-support structure enables the 2D color code to support transversal implementation of the full Clifford gate set, a significant advantage for fault-tolerant logical operations. Logical operators correspond to strings of $X$ or $Z$ operators spanning the lattice boundary-to-boundary; the triangular patch encodes a single logical qubit at any distance.
For the odd-distance triangular 6.6.6 color codes considered here, the number of data qubits is $N=(3d^2+1)/4$, with $N-1$ independent stabilizer generators. The $d=3$ instance is the $[[7,1,3]]$ Steane code, with 7 data qubits and 6 stabilizer generators. We also consider $d=5$ (19 data qubits and 18 stabilizer generators) and $d=7$ (37 data qubits and 36 stabilizer generators).

On the 6.6.6 lattice the bulk stabilizers have weight 6, with weight-4 stabilizers only at the boundary; the $d=3$ instance is the exception, being small enough that all three faces are boundary faces of weight 4. Compared with the weight-4 bulk stabilizers of the surface code, the higher stabilizer weight requires longer syndrome-extraction circuits and therefore introduces more gate-error locations per round, a tradeoff against the transversal gate-set advantage.

\subsection{Surface Code}

We consider the rotated surface code, in which data qubits and ancilla qubits are arranged on a two-dimensional lattice with nearest-neighbor interactions only. Stabilizers are of two types: X-type stabilizers detect Z errors, and Z-type stabilizers detect X errors, each acting on up to four neighboring data qubits (two at boundaries), with X- and Z-type stabilizers occupying separate plaquettes, unlike the color code's shared faces. Logical operators are strings of X or Z operators running edge-to-edge across the lattice; the minimal planar patch encodes a single logical qubit.
The code distance d equals the linear lattice size. A distance-d instance requires $N=d^2$ data qubits and N-1 ancilla qubits (2N-1 total). The $d=3$ instance, the smallest capable of demonstrating error correction, has 9 data qubits and 8 ancilla qubits (17 total); we also use $d=5$ (25 data, 24 ancilla, 49 total) and $d=7$ (49 data, 48 ancilla, 97 total). Its strictly local weight-4 (or boundary weight-2) stabilizers and high fault-tolerance threshold make the rotated surface code particularly well suited to planar qubit array platforms.

\section{LUT Decoding of Repetition Codes}
\label{sec:perf_lut}

\begin{figure}[!htbp]
\centering
\includegraphics[width=\columnwidth]{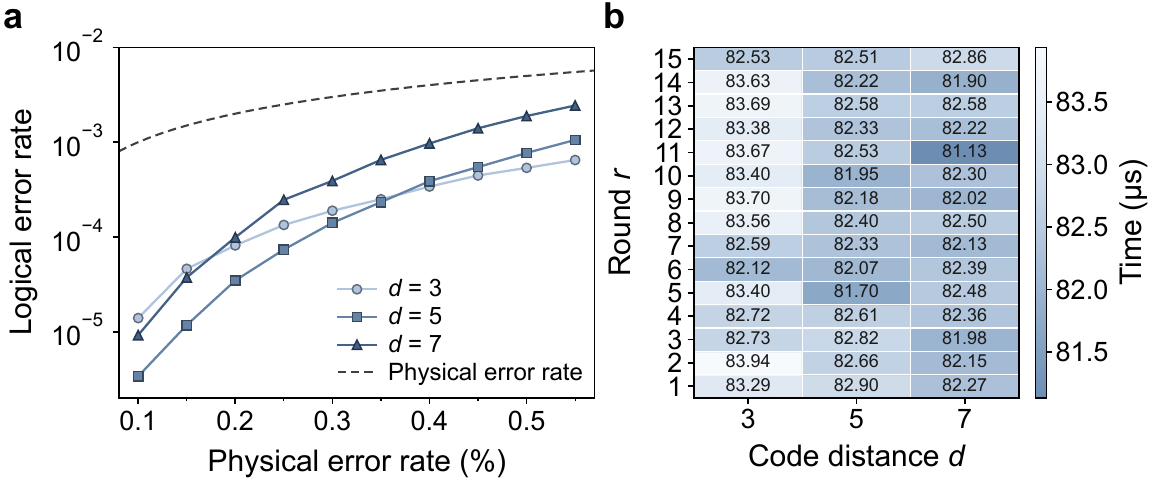}
\caption{\textbf{Performance of the LUT decoder for the repetition code.} (a) Logical error rate (LER) as a function of physical error rate (PER). The dashed line marks $\text{LER} = \text{PER}$. (b) GPU decoding time per shot for combinations of code distance $d$ and the number of syndrome-extraction rounds $r$ at $\text{PER} = 0.1\%$. Decoding times range from 81.1 to 83.9\,$\mu\mathrm{s}$. All reported values are averaged over $10^5$ shots.}
\label{fig:ler_rep}
\end{figure}

We employ lookup-table (LUT) decoding for repetition codes with distances $d=3,5,7$ as a low-latency reference. The LUT is constructed by enumerating error patterns of Hamming weight up to three, with the corresponding syndrome--correction pairs precomputed offline. Figure~\ref{fig:ler_rep}(a) shows the logical error rate (LER) as a function of physical error rate (PER). All three code distances achieve an LER below the reference line $p_L=p$ over the investigated range. At higher PER, however, the advantage of larger code distances diminishes and eventually reverses, reflecting the limited coverage of higher-weight error patterns by the truncated LUT.



Figure~\ref{fig:ler_rep}(b) shows the GPU decoding time as a function of code distance and syndrome-extraction rounds at a PER of $0.1\%$. The decoding time remains approximately constant at $82\,\mu$s across all tested configurations, indicating that the measured latency is dominated by fixed software overhead rather than the lookup operation itself. These results establish LUT decoding as a low-latency reference for the FPGA--GPU architecture, while its limited error-pattern coverage restricts its applicability to larger codes and higher physical error rates.

\section{AI Decoder: Training, Timing Structure, and Benchmark Settings}
\label{app:ai_decoder}

\paragraph{Training and transfer across distances.}
A single network configuration is used at all three code distances, and the model is transferred from small to large distance rather than trained from scratch at each $d$. All parameters are shared across stabilizers, so the code distance enters only as the length of the stabilizer axis and no weight tensor changes shape. Training uses detection-event records and the corresponding logical-observable flips sampled with Stim~\cite{gidney2021stim}, spanning a range of physical error rates and syndrome-extraction rounds up to $r = 35$.

\paragraph{Temporal compression and timing structure.}
Following the learned temporal compression of Ref.~\cite{senior2025scalable}, the decoder combines groups of $k = 6$ consecutive rounds into a single group representation. Within a processing block, the computations for all groups are parallelized across the spatial layers, while a recurrent state carries the syndrome history across blocks. The cumulative decoding time is therefore constant within a group and increases in discrete steps at each group boundary; each step comprises the transfer of the additional six rounds of syndrome data to the GPU, their preparation, and the corresponding inference. The trained network executes a fixed sequence of operations whose computational graph depends only on the architecture and the input dimensions. Varying the physical error rate therefore changes the syndrome values but not the amount of computation, so the decoding time is independent of the physical error rate by construction, in contrast to BP-OSD, whose runtime grows with the detection-event density.

\paragraph{Benchmark settings.}
The MWPM baseline~\cite{higgott2025sparse,dennis2002} maps each detection-event pattern onto a graph whose vertices are the triggered detectors and returns the pairing of lowest total weight, corresponding to the most likely set of errors consistent with the observed syndrome; no correlated reweighting is applied. The number of shots is adjusted for each parameter setting to reduce statistical uncertainty: $10^7$ shots for $d=7$ at $p=0.001$; $10^6$ shots for $d=5$ with $0.001\le p\le0.002$ and for $d=7$ with $0.001<p\le0.0025$; and $10^5$ shots for all other parameter settings.

\paragraph{Shot-to-shot variability.}
For a fixed code distance, number of syndrome-extraction rounds, and decoder configuration, the LUT and AI decoders have an approximately fixed computational workload per shot; residual timing variations arise from software and hardware effects such as scheduling and memory access. In contrast, the workload of the BP-OSD decoder depends on the input syndrome: different syndromes may require different numbers of BP iterations, and OSD post-processing is invoked only if BP fails to converge within the maximum number of iterations. These syndrome-dependent execution paths can lead to greater shot-to-shot variability in decoding time.

\section{Single-Round Logical Error Rate Extraction}
\label{app:sre}

A core metric of decoder performance is the logical error rate, declared for a logical qubit whenever its decoded state disagrees with the true state. Because errors accumulate over repeated stabilizer-measurement cycles, the cumulative logical error rate after $r$ rounds, assuming a constant single-round rate $p_L$ and independent per-round error events, follows
$$p_{\mathrm{cum}}(r) = \frac{1}{2}\left[1 - (1 - 2 p_L)^r\right],$$
estimated in practice by counting logical errors across many independent decoding instances. The corresponding decoding fidelity is $$F(r) = 1 - 2p_{\mathrm{cum}}(r) = (1-2p_L)^r,$$ whose logarithm $$\ln F(r) = r\ln(1-2p_L)$$ is linear in $r$ with zero intercept. Because state preparation and measurement (SPAM) introduce additional errors independent of $r$, we admit an offset $\omega$ and fit
$$\ln F(r) = \omega + \varepsilon r,$$
so that the single-round logical error rate follows from the slope as $$p_L = \frac{1}{2}\left(1 - e^{\varepsilon}\right).$$ At $r=0$ the fit gives $F(0)=e^{\omega}$, so $\omega$ absorbs the SPAM-induced offset and is negative for any nonzero preparation and measurement error. It does not enter $p_L$, which depends only on the slope. This turns the exponential growth of the cumulative error rate into a linear fitting problem, improving the stability of the extracted $p_L$. All figures in Section~\ref{sec:perf} (decoding results) report $p_L$ obtained by this fit.




\end{document}